# Nanoscale Quantification of Octahedral Tilts in Perovskite Films

Jinwoo Hwang[a)], Jack Y. Zhang, Junwoo Son, and Susanne Stemmer[b)]

Materials Department, University of California, Santa Barbara, California 93106-5050, USA

**Abstract**

NiO$_6$-octahedral tilts in ultrathin LaNiO$_3$ films were studied using position averaged convergent beam electron diffraction (PACBED) in scanning transmission electron microscopy. Both the type and magnitude of the octahedral tilts were determined by comparing PACBED experiments to frozen phonon multislice simulations. It is shown that the out-of-plane octahedral tilt of an epitaxial film under biaxial tensile stress (0.78 % in-plane tensile strain) increases by ~ 20%, while the in-plane rotation decreases by ~ 80%, compared to the unstrained bulk material.

[a] Electronic mail: jhwang@mrl.ucsb.edu

[b] Electronic mail: stemmer@mrl.ucsb.edu

Tilt patterns formed by the corner-sharing $BO_6$ octahedra in perovskite unit cells (general formula $ABO_3$) not only account for the wide range of structures and symmetries [1], but are also important in determining their properties. For example, the type and degree of $NiO_6$ octahedral tilts in the rare-earth nickelates ($R$NiO$_3$, where $R$ is a rare-earth cation) is correlated with the temperature of the metal-insulator transition [2]. Recent studies of thin film perovskites have shown that epitaxial strains modify octahedral tilts [3-5]. Epitaxial strains also change the properties such as $d$-band width, electrical resistivity and metal-insulator transitions of $R$NiO$_3$ films [6-8]. Thus, quantifying octahedral tilts as a function of film strain is essential for understanding the structure-property relationships of these materials. Prior studies have used x-ray diffraction to determine octahedral tilts in perovskite thin films [3,4,9]. Here, we present an alternative approach, position averaged convergent beam electron diffraction (PACBED), to quantify the type and magnitude of octahedral tilts in ultrathin perovskite films. PACBED involves spatially averaging coherent CBED patterns, formed using an atomic resolution scanning transmission electron microscopy (STEM) probe, over one or a few unit cells [10]. By comparing experimental patterns to simulations, PACBED can be used to accurately measure the thicknesses of TEM samples [10,11], and to detect subtle structural asymmetries, such as the direction of the polarization in ferroelectric films [12]. In contrast to x-ray diffraction, PACBED allows for nanoscale characterization in areas as small as a unit cell. PACBED uses the same electron optical configuration as high-angle annular dark-field (HAADF) imaging, so lattice resolution images can be acquired in parallel.

In the present study, PACBED patterns were obtained from a 5 nm thin LaNiO$_3$ film grown on a (001) (LaAlO$_3$)$_{0.3}$(Sr$_2$AlTaO$_6$)$_{0.7}$ (LSAT) substrate [6]. Bulk LaNiO$_3$ is rhombohedral (space group $R\bar{3}c$) with a $a^-a^-a^-$ type octahedral tilt pattern in Glazer notation and a tilt angle $\alpha = 5.2°$ [2]. In the Glazer notation [1], the letters *a*, *b*, and *c* indicate the relative magnitudes of the rotations about each of the three Cartesian axes, and the superscript indicates the relationship between two neighboring octahedra along each axis (+, -), or denotes no rotation along that axis (0). For example, $a^+a^-c^0$ means the same amount of rotation about *a* and *b* axes, no rotation about the *c* axis, and that two neighboring octahedra along the *a* axis rotate in the same direction while those along the *b* axis rotate in opposite directions. The angles of rotation about the *a*, *b*, and *c* axes are denoted $\alpha$, $\beta$, and $\gamma$, respectively. Films were capped with a 10 nm Ni layer for protection, and TEM samples were prepared by 2° wedge polishing. PACBED patterns and HAADF images were acquired in an FEI Titan STEM operated at 300 kV. The measured convergence angle of the STEM probe was 9.6 mrad, and PACBED patterns were acquired with an 8 second exposure on a charge coupled device camera.

Figure 1(a) shows a PACBED pattern acquired from the $LaNiO_3$ film area indicated by the box in the HAADF image in Fig. 1(b). Figure 1(c) is the same pattern with the Bragg disks labeled using pseudocubic indices. The most prominent features of the patterns arise from the overlapping 011 and 020 disks. The LSAT substrate induces an in-plane tensile strain (~ 0.78 %) in the $LaNiO_3$ film [6], causing the distance between 020 and $0\bar{2}0$ disks to be shorter than the distance between 002 and $00\bar{2}$ disks. Experimental PACBED patterns were compared to frozen phonon multislice simulations, which were carried out using the code by Kirkland [13]. The lattice parameters of the $LaNiO_3$ film used in the simulation were determined both by x-ray diffraction (XRD) and PACBED. The in-plane lattice parameter of epitaxial $LaNiO_3$ film was determined to be that of the LSAT substrate (0.387 nm), as expected for a fully strained film. The out-of-plane lattice parameter obtained from XRD was 0.382 nm [6]. Measuring the distances between the disks in PACBED patterns yielded a similar ratio of in- and out-of-plane lattice parameters (0.983 ± 0.004). Simulations were carried out for different types and amounts of octahedral rotations, using the Debye Waller factor of bulk LaNiO$_3$ [14]. The thickness of the TEM sample was determined by comparing experimental and simulated patterns using an automated algorithm (discussed below), after scaling the maximum and minimum intensities [10]. Figure 2 shows that the PACBED patterns are sensitive to both type and amount of octahedral rotation. A very wide range of tilt patterns and angles were explored (not shown). The best visual match to the experiment shown in Fig. 1(a) corresponds to a pattern with $a^-a^-c^-$ tilt, with $\alpha = 7°$ and $\gamma = 1°$, as shown in Fig. 2(b). This pattern shows greater intensity due to the overlapping 011 disks, compared to a pattern from the cubic unit cell with no octahedral tilts [Fig. 2(a)], which shows relatively stronger 020 disks. The pattern in Fig 2(b) is also clearly different from the pattern with larger angles in Fig. 2(c), where the intensities of the 011 disks are even stronger, or from the pattern with a different rotation type, $a^+a^+c^+$ in Fig. 2(d), which shows weaker 011 disks.

For quantitative comparisons of a large number of possible tilt patterns, an automated comparison algorithm was used, which calculates the minimum $\chi^2$ for the intensity difference between experimental and simulated PACBED patterns. The experimental PACBED patterns usually contain more pixels than the simulations, so they were first reduced to match the size of the simulated patterns. Then, $\chi^2$ is calculated as [13]:

$$\chi^2 = \frac{1}{N_x N_y} \sum_{i,j} \left[ I_{\exp}(x_i, y_j) - I_{\sim}(x_i, y_j) \right]^2, \qquad (1)$$

where $N_xN_y$ is the total number of pixels in the pattern, and $I(x_i, y_j)$ is the intensity at the pixel position $(x_i, y_j)$. The scattering angle was limited to 15.3 mrad in the calculation. Figure 3(a) shows a $\chi^2$ map of the comparison between the experimental PACBED pattern of a 13 nm thick region [shown in Fig. 3(b)] and simulated patterns of the $a^-a^-c^-$ tilt type, as a function of $\alpha$ and $\gamma$. Here, $a^- a^- c^-$ type patterns were simulated for 100 different tilt configurations, by varying $\alpha$ and $\gamma$ from 0 to 9°, for different foil thicknesses. The simulated patterns were compared to the experiment as a function of thickness, using Eq. (1). A thickness of 13 nm ± 1% gave the best match for all $\alpha$ and $\gamma$. The contour lines in Fig. 3(a) are fits to a two-dimensional polynomial function. The minimum of the fit function corresponded to $\alpha = 6.2 \pm 0.41°$ and $\gamma = 0.9 \pm 0.82°$, where the uncertainty values given here correspond to the errors in the fit. Thus, the automated comparisons resulted in a slightly smaller value for $\alpha$ than the match obtained by simple visual inspection. The map also shows that the PACBED patterns are more sensitive to $\alpha$ than to $\gamma$, as indicated by the errors stated above. The results are very similar to the results from an x-ray diffraction study of a thicker (17 nm) $LaNiO_3$ film on $SrTiO_3$ [4], which is also under tensile strain. In ref. [4], the in-plane rotation, $\gamma$, was $0.3 \pm 0.7°$, while the out-of-plane rotation, $\alpha$, was $7.1 \pm 0.2°$. Small differences in the in-plane rotations could be due to the differences in tensile strain in the present study and ref. [4].

The experimental pattern in Fig. 1(a) is asymmetric along the film normal, in particular with respect to the weak 013 and 0$\bar{1}$3 disks. This is likely caused by a slight deviation of the sample from the exact zone axis. Simulated patterns in Fig. 3 show that the 013 and 0$\bar{1}$3 disks disappear as the sample is tilted by 1 to 2 mrad away from the exact <100> zone axis along the *c* direction. However, even in the presence of a small sample tilt, Fig. 3 shows that the main characteristics sensitive to octahedral rotation, such as the overlapping intensity of 011 disks, do not change significantly as a result of sample tilt. In addition, the experimental pattern is less detailed than the simulations, which was also the case in the previous PACBED experiments [10,12]. Defects and amorphous layers on the surface decrease the contrast in the experimental pattern.

In summary, we have shown that PACBED can be used to determine octahedral tilts in ultrathin perovskite films despite large unit cells, which cause significant overlap of diffraction disks for the convergence angles used in HAADF imaging. The method allows for the investigation of the local structural origins that determine the properties of nanoscale heterostructures with distorted perovskites, including Mott materials such as the rare earth nickelates or titanates.

J. H., J. Y. Z. and S.S. acknowledge support from DOE (DEFG02-02ER45994) and NSF (Grant DMR-0804631). Partial support for J. H. was also provided by the MRSEC Program of the National Science Foundation (Award No. DMR 1121053). J. S. and S.S. were also supported through a grant by the Army Research Office (W911-NF-09-1-0398). This work made use of facilities supported by the Center for Scientific Computing at the California Nanosystems Institute (NSF CNS-0960316) and the UCSB Materials Research Laboratory, an NSF-funded MRSEC (DMR-1121053).

**References**

[1]     A. M. Glazer, Acta Cryst. B **28**, 3384 (1972).

[2]     J. B. Torrance, P. Lacorre, A. I. Nazzal, E. J. Ansaldo, and C. Niedermayer, Phys. Rev. B **45**, 8209 (1992).

[3]     A. Vailionis, H. Boschker, W. Siemons, E. P. Houwman, D. H. A. Blank, G. Rijnders, and G. Koster, Phys. Rev. B **83**, 064101 (2011).

[4]     S. J. May, J. W. Kim, J. M. Rondinelli, E. Karapetrova, N. A. Spaldin, A. Bhattacharya, and P. J. Ryan, Phys. Rev. B **82**, 014110 (2010).

[5]     J. C. Woicik, C. K. Xie, and B. O. Wells, J. Appl. Phys. **109**, 083519 (2011).

[6]     J. Son, P. Moetakef, J. M. LeBeau, D. Ouellette, L. Balents, S. J. Allen, and S. Stemmer, Appl. Phys. Lett. **96**, 062114 (2010).

[7]     G. Catalan, R. M. Bowman, and J. M. Gregg, Phys. Rev. B **62**, 7892 (2000).

[8]     J. A. Liu, M. Kareev, B. Gray, J. W. Kim, P. Ryan, B. Dabrowski, J. W. Freeland, and J. Chakhalian, Appl. Phys. Lett. **96**, 233110 (2010).

[9]     F. Z. He, B. O. Wells, Z. G. Ban, S. P. Alpay, S. Grenier, S. M. Shapiro, W. D. Si, A. Clark, and X. X. Xi, Phys. Rev. B **70**, 235405 (2004).

[10]    J. M. LeBeau, S. D. Findlay, L. J. Allen, and S. Stemmer, Ultramicroscopy **110**, 118 (2010).

[11]    R. Bjørge, C. Dwyer, M. Weyland, P. N. H. Nakashima, C. D. Marioara, S. J. Andersen, J. Etheridge, and R. Holmestad, Acta Mater. **60**, 3239 (2012).

[12]    J. M. LeBeau, A. J. D'Alfonso, N. J. Wright, L. J. Allen, and S. Stemmer, Appl. Phys. Lett. **98**, 052904 (2011).

[13] E. J. Kirkland, Advanced Computing in Electron Microscopy, 2nd ed. (Springer, New York, 2010).

[14] J. L. García-Muñoz, J. Rodríguez-Carvajal, P. Lacorre, and J. B. Torrance, Phys. Rev. B **46**, 4414 (1992).

**Figure captions**

**Figure 1 (color online):** (a) Experimental PACBED pattern acquired from the area outlined by the orange box shown in (b). (b) HAADF-STEM image of a 5 nm thick LaNiO$_3$ film on LSAT. (c) Same pattern as in (a), with pseudocubic indices.

**Figure 2 (color online):** Simulated PACBED patterns (right) for the structures shown on the left: (a) cubic LaNiO$_3$ with no octahedral tilts and (b-d) strained LaNiO$_3$ with (b) $a^-a^-c^-$ tilt with $\alpha = 7°$, $\gamma = 1°$, (c) $a^-a^-c^-$ with $\alpha = 9°$, $\gamma = 5°$, and (d) $a^+a^+c^+$ with $\alpha = 7°$, $\gamma = 1°$. All simulations are for a 11.6 nm TEM foil thickness. The film normal is parallel to $c$.

**Figure 3 (color online):** (a) $\chi^2$ map of the comparison between the experimental PACBED pattern [shown in (b)], and simulated patterns as a function of the octahedral tilt angles, $\alpha$ and $\gamma$, in the simulations. The TEM foil thickness is 13 nm. The contour lines are a 2D polynomial fit to the map. The $\chi^2$ values are normalized to the maximum pixel value of the map. The simulated pattern for the tilt angles that are closest to the minimum $\chi^2$ obtained from the fit function is shown in (c).

**Figure 4 (color online):** Simulated PACBED patterns of LaNiO$_3$ on LSAT with octahedral rotation $a^-a^-c^-$ with $\alpha = 7°$, $\gamma = 1°$, for TEM sample tilts that deviate by 0 to 2 mrad from the zone axis in the $c$ direction.

**Figure 1**

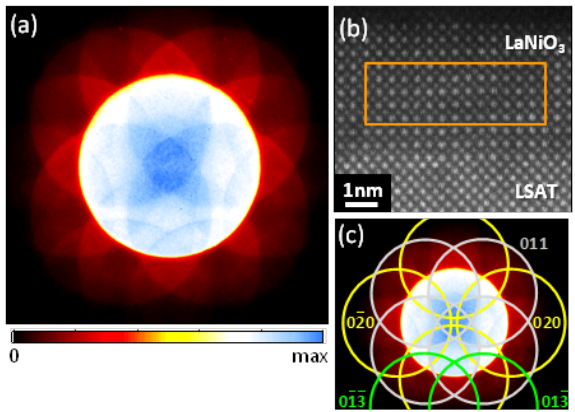

**Figure 2**

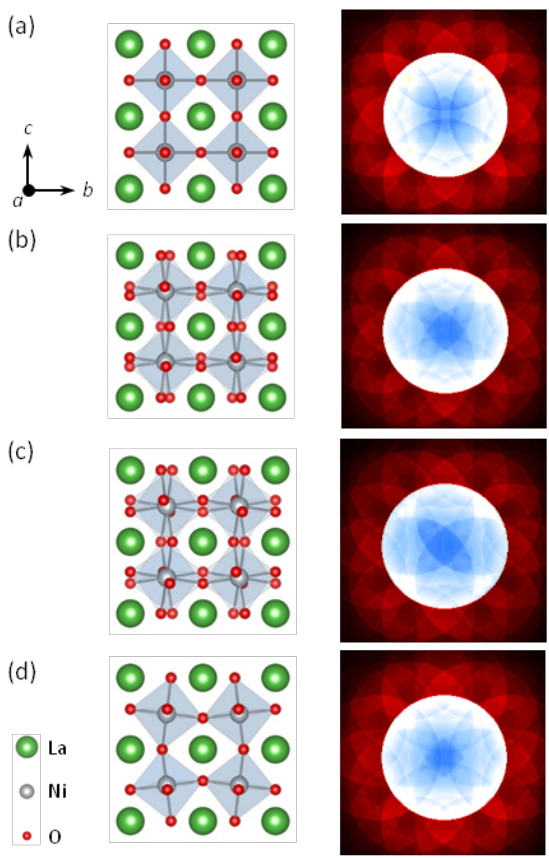

**Figure 3**

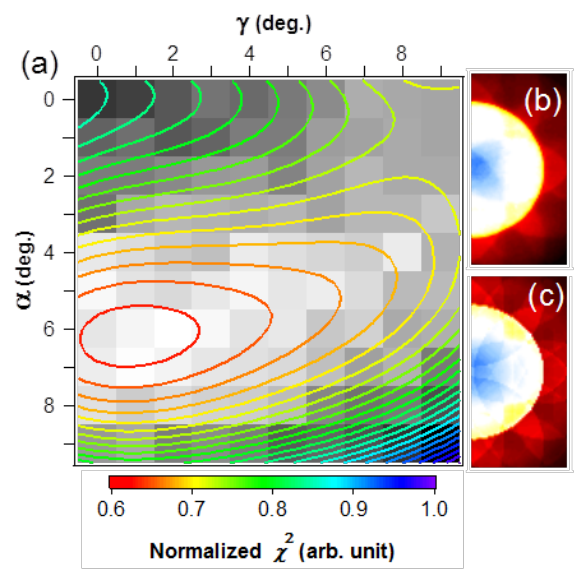

**Figure 4**

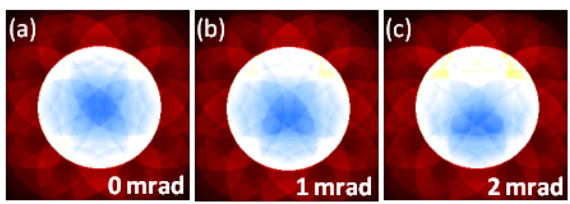